\documentstyle[12pt]{article}

\def\nqq{\hspace{-2em}}

\newcommand{\beq}[1]{\begin{equation}\label{#1}}
\newcommand{\eeq}{\end{equation}}
\newcommand{\bear}[1]{\begin{eqnarray}\label{#1}}
\newcommand{\ear}{\end{eqnarray}}

\def\barr{\left(\begin{array}}
\def\earr{\end{array}\right)}

\def\beq#1{\begin{equation}\label{#1}}
\def\eeq{\end{equation}}
\def\ber#1{\begin{eqnarray}\label{#1} \nqq}
\def\eer{\end{eqnarray}}

\def\mm{\\ \nqq}

\newcommand{\np}{ {\newpage } }

\newcommand{\N}{ \mbox{\rm I$\!$N} }
\newcommand{\R}{ \mbox{\rm I$\!$R} }
\def\C{\mbox{\rm {I\kern-.520em C}}}

\newcommand{\p}{\partial}
\newcommand{\btd}{\bigtriangledown}
\newcommand{\btu}{\bigtriangleup}
\newcommand{\tri}{\Delta}

\begin{document}
\centerline{\large \bf
P-brane Black Holes and
Post-Newtonian Approximation}

\vspace{1.03truecm}
\bigskip
\centerline{\bf \large  S. Cotsakis$^{\dagger}$\footnote{email:
skot@aegean.gr},
V. D. Ivashchuk$^{\ddagger}$\footnote{email: ivas@rgs.phys.msu.su}
 and V. N. Melnikov$^{\dagger}$\footnote{email:
melnikov@rgs.phys.msu.su}$^,$\footnote{On leave from $\ddagger$}}

\vspace{0.5cm}

\centerline{$\dagger$ Research Laboratory for Geometry, Dynamical
Systems and Cosmology
(GEODYSYC)}
\centerline{Department of Mathematics}
\centerline{University of the Aegean}
\centerline{Karlovassi, 83 200 Samos, Greece}

\vspace{0.2cm}

\centerline{$\ddagger$ Center for Gravitation and Fundamental
Metrology}
\centerline{VNIIMS, 3-1 M. Ulyanovoy Str.}
\centerline{Moscow, 117313, Russia}

\vspace{1cm}

\begin{abstract}
We analyze $p$-brane black hole solutions
with `block-orthogonal' intersection rules.
The post-Newtonian parameters  $\beta$ and $\gamma$
corresponding to  $4$-dimensional section
of the metric are calculated.
A family of solutions  with $\gamma =   1$
is singled out. Some examples of solutions
(e.g. in $D=  11$ supergravity) are considered.

\end{abstract}

\hspace*{0.950cm} PACS number(s): \ 04.50, \ 04.65, \ 98.80.H, \
04.60.Kz
\np

\section{\bf Introduction}
\setcounter{equation}{0}

Exact spherically symmetric solutions describing generalized
analogues of black holes in an arbitrary number of dimensions
have an interesting history and have recently received renewed
attention mainly in efforts to obtain a framework for a unified
theory, for example in the context of strings or $p$-branes.

The first  multidimensional generalization of such
type was considered by D. Kramer \cite{Kr} and rediscovered by A.I. Legkii
\cite{Le}, D.J. Gross and M.J. Perry  \cite{GP} (and also by A. Davidson
and D. Owen \cite{DO}, see also \cite{GW}).
In \cite{BrI} the Schwarzschild solution
was generalized to the case of $n$ internal Ricci-flat spaces and it was
shown that a black hole configuration takes place when the scale factors
of the internal spaces are constants. In \cite{FIM2}, an analogous
generalization of the Tangherlini solution \cite{Tan} was obtained. These
solutions were also extended to the electrovacuum  \cite{FIM3,IM8,BM} and
dilatonic \cite{BI,BM}  cases. (We remind that the
multidimensional $O(d+1)$-symmetric analogue of
the well-known  Reissner-Nordstr\"om charged black hole solution
was  obtained earlier by R. C. Myers and M. J. Perry \cite{MP}.)
A theorem was proved
in \cite{BM} that cuts  all non-black-hole configurations as
non-stable under even monopole perturbations.

 In \cite{IM13} the
extremely-charged dilatonic black hole solution was generalized
to the multicenter (Majumdar-Papapetrou) case when the
cosmological constant is non-zero.
(The $D =  4$   Majumdar-Papapetrou solutions with conformal
scalar  and electromagnetic fields were considered already in
\cite{Br}.) In \cite{IMC,IMBl}, the Majumdar-Papapetrou type
solutions with composite intersecting $p$-branes (in theories
with fields of forms) \footnote{For non-composite electric
case see \cite{IM0,IM}, for composite electric case see
\cite{IMR}, for solutions with intersections governed by Lie
algebras see \cite{GrI}. For other solutions with $p$-branes
see also  \cite{DGHR}-\cite{St} and references therein.}
corresponding to Ricci--flat internal spaces were obtained
and generalized to the case of Einstein internal spaces.
Earlier some special classes of these solutions were considered
in \cite{PT}--\cite{AIR}. The obtained solutions take place when
certain
orthogonality relations (on coupling parameters, dimensions of
branes, total dimension) are imposed. In such a situation, one
may have a new class of cosmological and spherically symmetric
solutions \cite{IMJ}. Special cases were
also considered in \cite{LPX}-\cite{BKR}.
Solutions with a horizon were studied in detail in
\cite{CT}-\cite{BIM}, \cite{IMJ}. In \cite{BIM,Br1} some propositions
related to i) the interconnection between  Hawking temperature
and  singularity behaviour and ii) multitemporal
configurations were proved. It should be noted that the
multidimensional and multitemporal  generalizations of the
Schwarz\-schild and Tangherlini  solutions were considered in
\cite{IM8,IM6I} wherein  generalized  Newton's formulas for the
multitemporal case were obtained.

The plan of this letter is as follows. In Section 2,  we consider
$p$-brane black hole (BH) solutions
with `block-orthogonal' intersection
rules  \cite{IMBl} (see also \cite{Br1}) and provide an  example of a
 black hole  solution in $D=   11$ supergravity \cite{CJS}.
The metric of this solution contains the Reissner-Nordstr\"om
 metric as a 4-dimensional section.
In Section 3, the post-Newtonian parameters  $\beta$ and $\gamma$
corresponding to  $4$-dimensional section
of the metric are calculated and a
 family of solutions  with $\gamma =   1$,
corresponding to electro-magnetic pairs of $p$-branes
is singled out. Some comments of a more general nature are given in the
last Section.

\section{\bf $p$-brane black holes}
\setcounter{equation}{0}

Our starting point is the action
\ber{2.1}
S =
\int_{M} d^{D}z \sqrt{|g|} \{ {R}[g] - h_{\alpha\beta}
g^{MN} \partial_{M} \varphi^\alpha \partial_{N} \varphi^\beta
- \sum_{a \in \Delta}
\frac{1}{n_a!} \exp[ 2 \lambda_{a} (\varphi) ] (F^a)^2_g \},
\eer
where $g =   g_{MN} dz^{M} \otimes dz^{N}$ is the metric
of signature $(-,+, \ldots, +)$ ($M,N =  1, \ldots, D$),
$\varphi=  (\varphi^\alpha)\in \R^l$
is a vector from dilatonic scalar fields,
$(h_{\alpha\beta})$ is a non-degenerate
symmetric $l\times l$ matrix ($l\in \N$),
\beq{2.2}
F^a =    dA^a
=  \frac{1}{n_a!} F^a_{M_1 \ldots M_{n_a}}
dz^{M_1} \wedge \ldots \wedge dz^{M_{n_a}}
\eeq
is a $n_a$-form ($n_a \geq 2$) on a $D$-dimensional manifold $M$, and
$\lambda_{a}$ is a $1$-form on $\R^l$ :
$\lambda_{a} (\varphi) =  \lambda_{a \alpha}\varphi^\alpha$, $a \in
\Delta$, $\alpha=  1,\ldots,l$.
In (\ref{2.1})
we denote $|g| =   |\det (g_{MN})|$,
\beq{2.3}
(F^a)^2_g  =
F^a_{M_1 \ldots M_{n_a}} F^a_{N_1 \ldots N_{n_a}}
g^{M_1 N_1} \ldots g^{M_{n_a} N_{n_a}},
\eeq
$a \in \Delta$, where $\Delta$ is some finite set. Varying this
action with respect to $g$, $\varphi$ and $A^a$ we obtain the
equations of motion in  the
form,
\bear{2.4}
R_{MN} - \frac{1}{2} g_{MN} R  =     T_{MN},
\\
\label{2.5}
{\btu}[g] \varphi^\alpha -
\sum_{a \in \Delta}   \frac{\lambda^{\alpha}_a}{n_a!}
e^{2 \lambda_{a}(\varphi)} (F^a)^2_g =   0,
\\
\label{2.6}
\nabla_{M_1}[g] (e^{2 \lambda_{a}(\varphi)}
F^{a, M_1 \ldots M_{n_a}})  =    0,
\ear
$a \in \Delta$; $\alpha=  1,\ldots,l$.
Here, $\lambda^{\alpha}_{a} =   h^{\alpha \beta}
\lambda_{\beta a}$, where $(h^{\alpha \beta})$
is a matrix inverse to $(h_{\alpha \beta})$,
\bear{2.7}
T_{MN} =     T_{MN}[\varphi,g]
+ \sum_{a\in\Delta}  e^{2 \lambda_{a}(\varphi)} T_{MN}[F^a,g],
\ear
where,
\bear{2.8}
T_{MN}[\varphi,g]  =
h_{\alpha\beta}\left(\p_{M} \varphi^\alpha \p_{N} \varphi^\beta -
\frac{1}{2} g_{MN} \p_{P} \varphi^\alpha \p^{P} \varphi^\beta\right),
\\
T_{MN}[F^a,g] =   \frac{1}{n_{a}!}  [ - \frac{1}{2} g_{MN}
(F^{a})^{2}_{g} +
n_{a}  F^{a}_{M M_2 \ldots M_{n_a}} F_{N}^{a, M_2 \ldots
M_{n_a}}], \label{2.9}
\ear
and ${\btu}[g]$, ${\btd}[g]$
are the Laplace-Beltrami and covariant derivative operators
respectively corresponding to  $g$.
In the following we shall be interested in
 p-brane black hole  solutions
to the equations (\ref{2.4})-(\ref{2.9}) \cite{IMBl},
defined on the manifold
\beq{3.1}
M =    (R_{0}, +\infty )  \times
S^{d_0}  \times \R \times M_2 \times  \ldots \times M_{n}, \eeq where
 $R_0 > 0$ and $S^{d_0}$ is the $d_0$-dimensional unit sphere.
Black hole (BH) solution
may be obtained also from general spherically symmetric solutions
\cite{IMJ1}. In explicit form BH solution reads
\bear{3.2}
&&g=   U \biggl\{ \frac{dR \otimes dR}{1 - 2\mu / R^{\bar d}}  +
R^2  d \hat{\Omega}^2_{d_0} -
U_1 \left(1 - \frac{2\mu}{R^{\bar d }} \right)  dt \otimes dt
+ \sum_{i =   2}^{n} U_i \hat{g}^i  \biggr\},
\\
\label{3.3}
&&U =   \prod_{s \in S} H_s^{2 d(I_s) \nu_s^2/(D-2)},
\\
\label{3.4}
&&U_1 =   \prod_{s \in S} H_s^{-2 \nu_s^2},
\\
\label{3.5}
&&U_i =   \prod_{s \in S} H_s^{-2 \nu_s^2 \delta_{iI_s}},
\qquad i > 1,
\\
\label{3.6}
&&\varphi^\alpha =
\sum_{s \in S} \nu_s^2 \chi_s \lambda_{a_s}^\alpha
\ln H_s,
\\
\label{3.7}
&&F^a=   \sum_{s\in S_e} \delta^a_{a_s} d\Phi^s \wedge \tau(I_s) +
\sum_{s\in S_m} \delta^a_{a_s} e^{-2\lambda_a(\varphi)}*[d\Phi^{s}
\wedge\tau(I_s)].
\eer
Here, we have set $\bar{d} =   d_0 -1$, $R_0^{\bar{d}} =   2\mu$,
$a \in \Delta$, $\alpha=  1,\ldots,l$, and
\bear{3.8}
&&\Phi^s =   \frac{\nu_s}{H_s^{'  } },
\\
\label{3.9}
&&H_s =   1 + \frac{ P_s}{R^{\bar{d}}},
\\ \label{3.10}
&&H'_s=   \Bigl(1 - \frac{ P'_s}{H_s R^{\bar{d}}} \Bigr)^{-1}=
1 + \frac{ P'_s}{ R^{\bar{d}} +  P_s -  P_s^{'} }, \\
\label{3.11}
&& (P'_s)^2 =   P_s ( P_s + 2 \mu),
\ear
where $P_s > 0$ are constants, $s \in S$.
In  (\ref{3.2}),
$g^i  =   g^{i}_{m_{i} n_{i}}(y_i) dy_i^{m_{i}} \otimes
dy_i^{n_{i}}$ is the
 Euclidean Ricci-flat metric on $M_{i}$,
$i=  2,\ldots,n$ and $\hat{g}^{i} =   p_{i}^{*} g^{i}$ is the
pullback of  $g^{i}$  to the  $M$ by the
canonical projection: $p_{i} : M \rightarrow  M_{i}$, $i =   0,2,
\ldots, n$; $g^0 =   d \Omega^2_{d_0}$ is canonical
metric on $M_0 =   S^{d_0}$ and in (\ref{3.7})
$*=  *[g]$ denotes the Hodge operator on $(M,g)$.

The set $S$ (generalized $p$-brane set) is by definition
\ber{3.12}
S=  S_e \sqcup S_m, \quad
S_v=  \sqcup_{a\in\tri}\{a\}\times\{v\}\times\Omega_{a,v},
\eer
$v=  e,m$ where $\sqcup$ means the union of non-intersecting
sets and $\Omega_{a,e}, \Omega_{a,m} \subset \Omega$,
where $\Omega =   \Omega(n)$  is the set of all non-empty
subsets of $\{ 1, \ldots,n \}$. Thus any $s \in S$
has the form
\ber{3.13}
s =   (a_s,v_s, I_s),
\eer
where
$a_s \in \tri$, $v_s =  e,m$ and $I_s \in \Omega_{a_s,v_s}$.
The sets $S_e$ and $S_m$ define electric and magnetic $p$-branes. In
(\ref{3.6})
\ber{3.13c}
\chi_s  =   +1, -1
\eer
for $s \in S_e, S_m$ respectively.
In (\ref{3.5})
\beq{3.13i}
\delta_{iI}=  \sum_{j\in I} \delta_{ij}
\eeq
is the indicator of $i$ belonging
to $I$: $\delta_{iI}=  1$ for $i\in I$ and $\delta_{iI}=  0$ otherwise.

All the  manifolds $M_{i}$ are assumed to be oriented and
connected and  the volume $d_i$-forms
\beq{3.14}
\tau_i  \equiv \sqrt{|g^i(y_i)|}
\ dy_i^{1} \wedge \ldots \wedge dy_i^{d_i},
\eeq
are well--defined for all $i=  1,\ldots,n$.
Here $d_{i} =   {\rm dim} M_{i}$, $i =   0, \ldots, n$ ($M_0 =
S^{d_0}$), $D =   1 + \sum_{i =   0}^{n} d_{i}$, and for any
 $I =   \{ i_1, \ldots, i_k \} \in \Omega$, $i_1 < \ldots < i_k$,
we denote
\ber{3.15}
\tau(I) \equiv \tau_{i_1}  \wedge \ldots \wedge \tau_{i_k},
\\
\label{3.16}
M_{I} \equiv M_{i_1}  \times  \ldots \times M_{i_k},
\\
\label{3.17}
d(I) \equiv  \sum_{i \in I} d_i.
\eer
In our solution, since
\beq{3.18}
1 \in I_s
\eeq
for all $s \in S$ all     brane    manifolds $M_I$
contain the time submanifold $M_1 =   \R$.
Due to (\ref{3.7}), the dimension of
$p$-brane worldsheet $d(I_s)$ is defined by
\ber{3.19}
d(I_s)=  n_{a_s}-1, \quad d(I_s)=   D- n_{a_s} -1,
\eer
for $s \in S_e, S_m$ respectively.
(For a $p$-brane: $p =   p_s =   d(I_s)-1$).
The parameters  $\nu_s$ appearing in the solution
satisfy the relations
\ber{3.20}
\sum_{s \in S} B^{ss'} \nu_{s'}^2 =   1,
\eer
    with
\ber{3.21}
B^{ss'} \equiv
d(I_s\cap I_{s'})+\frac{d(I_s)d(I_{s'})}{2-D}+
\chi_s\chi_{s'}\lambda_{\alpha a_s}\lambda_{\beta a_{s'}}
h^{\alpha\beta},  \eer
and we assume that
\ber{3.22}
S=  S_1\sqcup\dots\sqcup S_k,
\eer
$S_i\ne\emptyset$, $i=  1,\dots,k$, and
\ber{3.23}
B^{ss'} =   0,
\eer
for all $s\in S_i$, $s'\in S_j$, $i\ne j$; $i,j=  1,\dots,k$.
Eq. (\ref{3.22}) means that the set $S$ is a union of $k$
non-intersecting (non-empty) subsets $S_1,\dots,S_k$ and
according to (\ref{3.23}) the matrix $B =   (B^{ss'})$ (\ref{3.21})
has a block-diagonal structure and is the  direct sum of $k$
blocks  $B^{(i)} =   (B^{ss'},s,s' \in S_i)$, $i=  1,\dots,k$, =
i.e.
\ber{3.24}
B =   {\rm diag}(B^{(1)}, \ldots, B^{(k)}).
\eer
(It is tacitly assumed that $S$ is ordered,
$S_1 < \ldots < S_k$, and the order in $S_i$ is inherited
by the order in $S$.)
The parameters $P_s$ coincide inside
blocks,
\ber{3.25}
P_{s} =   P_{s'},
\eer
for all $s,s' \in S_i$, $i=  1,\dots,k$.

The solution given above describes non-extremal charged
intersecting generalized  $p$-branes with block-diagonal
matrix $B$ and agrees in some particular cases with those given
in Refs. \cite{CT,AIV,Oh}
($d_1 =   \ldots =   d_n =  1$), \cite{BIM} for the
non-composite case and \cite{IMJ} for the case of diagonal $B$
(i.e. when $|S_1| =   \ldots =   |S_k| =   1$).
Note also that the special case of our solution with the
parameters $\nu_s^2$ coinciding inside blocks
(i.e. $\nu_{s}^2 =   \nu_{s'}^2$ for all $s,s' \in S_i$,
$i=  1,\dots,k$) was analyzed also in \cite{Br1}.

We now note that the metric (\ref{3.2}) has a horizon at
$R^{\bar{d}} =   2 \mu$.
The Hawking temperature corresponding to
the solution is (see also \cite{Oh,BIM} for     orthogonal    case)
found to be
 \beq{3.26}
T_H=   \frac{\bar{d}}{4 \pi (2 \mu)^{1/\bar{d}}}
\prod_{s \in S} \left(\frac{2 \mu}{2 \mu + P_s}\right)^{ \nu_s^2}.
\eeq

Therefore, for fixed $P_s > 0$
and $\mu \to + 0$, we deduce
$T_H(\mu) \to 0$ for the extremal
black hole configurations \cite{IMBl} satisfying
\beq{3.27}
\xi =   \sum_{s\in S} \nu_s^2- \bar{d}^{-1} > 0.
\eeq

{\bf Remark.}
Relation  (\ref{3.26}) may be obtained
using e.g. formulae from \cite{Wald}. One finds
for static metrics written in the form
\beq{3.28}
g = -\exp[2\gamma(u)]dt \otimes dt + \exp[2\alpha(u)]
du \otimes du  + \ldots
\eeq
the following expression for the Hawking temperature of a surface
$u=u^*$ where $\exp(\gamma)= 0$, assumed to be a horizon:
\beq{3.29}
T_H = \frac{1}{2\pi}\ \lim_{u\to u^*} \ \exp(\gamma-\alpha)
           \left|\frac{d\gamma}{du}\right|.
\eeq

It is instructive to give some examples that show the behaviour
of our solution in some simple cases.

{\bf Example 1: Reissner-Nordstr\"om solution.} Our solutions
contain the Reissner-Nordstr\"om solution (for a charged black
hole) as a special case, namely,
 \ber{3b.1} g=  H^2 \biggl\{
\frac{dR \otimes dR}{1 - 2\mu / R} + R^2  d \hat{\Omega}^2_{2}
\biggr\} - H^{-2} \left(1 - \frac{2\mu}{R} \right) dt\otimes dt
\\
\label{3b.2}
F =   \nu d(H^{'  })^{-1} \wedge dt
\eer
where $\nu^2 =   2$ and
\ber{3b.3}
H =   1 + \frac{P}{R}, \qquad H'=
1 + \frac{ P'}{ R +  P -  P' },
\eer
$(P')^2 =   P ( P + 2 \mu)$
and $P > 0$ is constant.  Introducing the new
radial variable $r =   R + P$, we  may rewrite
this solution in a more familiar form, namely,
\ber{3b.6}
g=   - f dt\otimes dt + r^2  d \hat{\Omega}^2_{2} +
f^{-1} dr \otimes dr,
\\
\label{3b.7}
F =   \nu \frac{P^{'  }}{r^2} dt \wedge dr,
\eer
where $\nu^2 =   2$,
\ber{3b.8}
f =   1 - \frac{2GM}{r} + \frac{(P^{'  })^2}{r^2},
\eer
with $GM =   \mu + P$.  (Here $G$ is gravitational constant,
$M$ is mass and $P^{'}$ is charge.)

{\bf Example 2: D=   11 supergravity.}
Consider the     truncated    bosonic sector of
$D=   11$ supergravity (    truncated    means without
Chern-Simons terms).  The action  (\ref{2.1}) in this
case reads
\ber{3a.1}
S_{tr} =   \int_{M} d^{11}z \sqrt{|g|} \{ {R}[g] - \frac{1}{4!}  F^2 \}.
\eer
where ${\rm rank} F =   4$. In this particular case,
 consider the dyonic black-hole solutions
with  electric 2-brane and magnetic  5-brane
defined on the manifold
\beq{3a.2a}
M =    (2\mu, +\infty )  \times
S^{2}  \times \R \times M_{2} \times M_{3},
\eeq
where ${\dim } M_2 =  2$ and ${\dim } M_3 =  5$.
The metric and 4-form field then  read as follows,
\ber{3a.2}
g=  H^2 \biggl\{ \frac{dR \otimes dR}{1 - 2\mu / R} +
R^2  d \hat{\Omega}^2_{2} \biggr\} -
H^{-2} \left(1 - \frac{2\mu}{R} \right) dt\otimes dt
+ \hat{g}^2 + \hat{g}^3, \mm
\label{3a.3}
F =   \nu_1 d(H^{'  })^{-1} \wedge dt\wedge \tau_2+
\nu_2 *(d (H^{'  })^{-1} \wedge dt \wedge \tau_3),
\eer
where $\nu_1^2 =   \nu_2^2 =  1$ the metrics $g^2$ and  $g^3$ are
Ricci-flat and the functions $H$ and $H^{'  }$ are defined by
(\ref{3b.3}).

The solution (\ref{3a.2}), (\ref{3a.3})
satisfies not only equations of motion for the truncated model
, but also  the equations of motion
for  $D =11$ supergravity with the bosonic sector action
\ber{3a.4}
S =  S_{tr} +  c \int_{M} A \wedge F \wedge F
\eer
($c = {\rm const}$,  $F = d A$),
since the only modification
related to "Maxwells" equations
\ber{6.15}
d*F = {\rm const} \ F \wedge F,
\eer
is trivial due to $F \wedge F = 0$ (since $\tau_i \wedge \tau_i =0$).

This solution describes two $p$-branes
(electric 2-brane and magnetic  5-brane)
with equal charges intersecting on the time manifold.
In the extremal case, $\mu \to 0$, this solution corresponds to
the so-called $A_2$-dyon solution  from \cite{IMBl}. We see that
the 4-dimensional section of the metric (\ref{3a.2}) coincides
with the Reissner-Nordstr\"om  metric (\ref{3b.1}).

\section{Post-Newtonian approximation}
Let $d_0 =   2$ and consider the 4-dimensional section of the
metric (\ref{3.2}), namely,
  \ber{4.1}
g^{(4)} =   U \biggl\{ \frac{dR \otimes dR}{1 - 2\mu / R} +
R^2  d \hat{\Omega}^2_{2} -
U_1 \left(1 - \frac{2\mu}{R} \right)  dt \otimes dt \biggr\}, \eer
in the range $R > 2\mu$.
We imagine that some real astrophysical objects (e.g. stars) may
be described by the 4-dimensional     physical    metric (\ref{4.1}),
i.e. they are     traces    of extended multidimensional objects
(charged $p$-branes). Introducing a new radial variable $\rho$ by
the relation \ber{4.2}
R =   \rho \left(1 + \frac{\mu}{2\rho}\right)^2,
\eer
($\rho > \mu/2$), we may rewrite the metric (\ref{4.1})
in the 3-dimensional conformally-flat form,
\ber{4.3}
g^{(4)} =   U \Biggl\{ -
U_1 \frac{\left(1 - \frac{\mu}{2 \rho} \right)^2}
{\left(1 - \frac{\mu}{2 \rho} \right)^2} dt \otimes dt +
\left(1 + \frac{\mu}{2 \rho} \right)^4
\delta_{ij} dx^i \otimes dx^j \Biggr\},
\eer
where $\rho^2 =  |x|^2 =   \delta_{ij}x^i x^j$ ($i,j =   1,2,3$).

For possible physical applications, one  should calculate the
post-Newtonian parameters $\beta$ and $\gamma$ (Eddington
parameters) using the following standard relations
\ber{4.4}
g^{(4)}_{00} =   - (1 -  2 V + 2 \beta V^2 ) + O(V^3),
\\
\label{4.5}
g^{(4)}_{ij} =   \delta_{ij}(1 + 2 \gamma V ) + O(V^2),
\eer
$i,j =   1,2,3$, where,
\ber{4.6}
V =   \frac{GM}{\rho}
\eer
is  Newton's potential, $G$ is the gravitational constant and
$M$ is the gravitational mass. From (\ref{4.3})-(\ref{4.6}) we
deduce the formulas
 \ber{4.7}
GM =   \mu + \sum_{s \in S}  \nu_s^2 P_s
\left(1 -  \frac{d(I_s)}{D-2} \right)
\eer
and
\ber{4.8}
\beta - 1 =   \frac{1}{2(GM)^2} \sum_{s \in S}  \nu_s^2 (P_s^{'})^2
\left(1 -  \frac{d(I_s)}{D-2} \right) \\
\label{4.9}
\gamma - 1 =   - \frac{1}{GM} \sum_{s \in S} \nu_s^2 P_s \left(1 -  2
\frac{d(I_s)}{D-2} \right).
\eer
The parameter $\beta$ is defined by the charges $P_s^{'}$ of
$p$-branes (or more correctly by the charge densities). It
follows from (\ref{4.8}) and the inequalities  $d(I_s) < D - 2$
(for all $s \in S$) that
\ber{4.10}
\beta \geq 1,
\eer
and $\beta =   1$ (as for the Schwarzschild solution)
only if all $P_s^{'} =   0$, i.e. in the pure vacuum case. As an
example, for the Reissner-Nordstr\"om solution
(\ref{3b.6}) we deduce,
\ber{4.11}
\beta =   1  + \frac{(P^{'})^2 }{2(GM)^2},
\\
\label{4.12}
\gamma =   1.
\eer
 Examples 1 and 2 imply that the same
parameters appear in the 4-dimensional section
of the $D=  11$ supergravity solution.

The results obtained above suggest that there exist non-trivial
p-brane configurations with $\gamma =  1$.  That this is indeed the
case is shown by the following

{\bf Proposition.}
{\em Let the set of $p$-branes consist of
several pairs of electric and magnetic branes and let
any such pair $(s, \bar{s} \in S)$  correspond to the same colour
index,
i.e. $a_s =   a_{\bar{s}}$ and the charge parameters are
equal, i.e. $P_s =   P_{\bar{s}}$. Then,
\ber{4.13}
\gamma =   1.
\eer  }

The Proposition can be readily proved using
the relation $d(I_s) + d(I_{\bar{s}}) =   D - 2$ following from
(\ref{3.19}).

Further, for  small $P_s$  ($P_s \ll \mu$) we obtain
\ber{4.16}
P_s \sim (P'_s)^2/(2\mu), \qquad GM \sim \mu
\eer
and hence
\ber{4.17}
\gamma - 1 \sim - \frac{1}{2(GM)^2} \sum_{s \in S}  \nu_s^2
(P_s^{'})^2 \left(1 -  2 \frac{d(I_s)}{D-2} \right). \eer
As a last example of our procedure,
let us consider the special  case of one $p$-brane.
Here we have,
\ber{4.18}
\nu^{-2}_s =   d(I_s) \left(1 -  \frac{d(I_s)}{D-2} \right) +
\lambda^2_{a_s}, \eer
and with the help of Eqs. (\ref{4.8}),  (\ref{4.9})  and
(\ref{4.18}) we see that for large enough values of
the dilatonic coupling constant it is possible to perform
a `fine tuning'  of the parameters $(\beta, \gamma)$ near
the point (1,1) even if the `charges'  $P_s^{'}$ are big.

\section{\bf Conclusions}
In this Letter, we analyzed the basic features of $p$--brane black hole
solutions and described some of the better known black hole spacetimes
as
limiting cases of $p$--brane black holes. A  restriction to the
$4$--dimensional sector of these metrics allowed the determination of
the
post--Newtonian parameters $\beta , \gamma$  and in particular, it was
shown
how a new family of solutions having $\gamma =  1$ could be singled
out.

Our results allow a number of more general comments to be made.
Firstly, there
is the question of the stability of our $p$--brane black holes under
small
perturbations away from our static configurations. It is known that in
the
multidimensional case, the class
of metrics which are stable against monopole perturbations is of
measure zero
in the space of all possible spherically symmetric solutions \cite{BM}.
 Therefore a more
general problem of stability of multidimensional black holes
involving other types of perturbations has to be
formulated and studied.

Secondly, it would be of interest to develop more general aspects of
$p$--brane
black holes addressing issues such as the area theorem and the laws of
$p$--brane black hole thermodynamics in this generalized context. Does
the
nondecreasing of area in the four dimensional sector imply
an area theorem for the
full multidimensional case? We leave such matters for the future.

\begin{center}
{\bf Acknowledgments}
\end{center}

The work of V.D.I and V.N.M was partially supported by the
  Russian Foundation for Basic Research grant `98-02-16414'
 and the project SEE.
 V.N.M. is grateful to the Department of Mathematics, University of
Aegean
 for their kind hospitality during his stay there.

\end{document}